\documentclass [12pt] {report}
\usepackage[dvips]{graphicx}
\usepackage{epsfig, latexsym,graphicx}
\usepackage{amssymb}
\newcommand {\D}[2] {\displaystyle\frac{\partial{#1}}{\partial{#2}}}

\newcommand {\Dd}[3] {\displaystyle\frac{\partial^2{#1}}{\partial{#2}\partial{#3}}}

\newcommand {\si} {\sigma}

\newcommand {\de} {\delta}

\newcommand {\fr} {\displaystyle\frac}

\newcommand {\be} {\begin{equation}}
\newcommand {\ee} {\end{equation}}
\newcommand {\ba} {\begin{array}}
\newcommand {\ea} {\end{array}}
\newcommand {\bp} {\begin{picture}}
\newcommand {\ep} {\end{picture}}
\newcommand {\bc} {\begin{center}}
\newcommand {\ec} {\end{center}}
\newcommand {\bt} {\begin{tabular}}
\newcommand {\et} {\end{tabular}}
\newcommand {\lf} {\left}
\newcommand {\rg} {\right}

\newcommand {\cF} {{\cal F}}

\newcommand {\cR} {{\cal R}}

\newcommand {\ses} {\medskip}

\newcommand {\bibit} {\bibitem}
\newcommand {\nin} {\noindent}

\newcommand {\De} {\Delta}

\usepackage{psfrag}

\usepackage{epsfig, latexsym,graphicx}
 \usepackage{amsmath}

\def\2#1#2#3{{#1}_{#2}\hspace{0pt}^{#3}}
\def\3#1#2#3#4{{#1}_{#2}\hspace{0pt}^{#3}\hspace{0pt}_{#4}}
\newcounter{sctn}
\def\sec#1.#2\par{\setcounter{sctn}{#1}\setcounter{equation}{0}
                  \noindent{\bf\boldmath#1.#2}\bigskip\par}

\begin {document}

\begin {titlepage}

\vspace{0.1in}

\begin{center}

{\Large \bf  Finsleroid-regular    space:~ curvature tensor, \\

continuation of gravitational Schwarzschild metric}

\end{center}

\vspace{0.3in}

\begin{center}

\vspace{.15in} {\large G.S. Asanov\\} \vspace{.25in}
{\it Division of Theoretical Physics, Moscow State University\\
119992 Moscow, Russia\\
{\rm (}e-mail: asanov@newmail.ru{\rm )}} \vspace{.05in}

\end{center}

\begin{abstract}

\ses

The method of simple straightforward calculation of the curvature tensor of the Finsleroid--regular space
is indicated.
The  Schwarzschild metric which underlines the  gravitational field produced by static spherical-symmetric body
is shown to be uniquely extended to the Finslerian domain upon a consistent treatment of
the pseudo-Finsleroid axis  vector field $b_i$ to be the field of the time variable.

\ses

\nin{\it Keywords:} Finsler  metrics, gravitational equations,
curvature tensors.

\end{abstract}

\end{titlepage}

\vskip 1cm

\ses

\ses

\setcounter{sctn}{1} \setcounter{equation}{0}

\nin
  {\bf 1. Description  of new conclusions}

\ses

\ses

The regular Finsleroid-Finsler spaces [1]
of the positive-definite type
 $\mathbf\cF\cR^{PD}_{g;c}$
 and of the relativistic case
  $\mathbf\cF\cR^{SR}_{g;c} $
are constructed with the help of the set
$\{g(x), b_i(x), a_{ij}(x),y\}$,
where
$g(x)$ is a scalar which plays the role of the Finsleroid charge,
$b_i(x)$ is an involved vector field, and $a_{ij}(x)$ is a metric tensor (which is positive-definite
in the space
 $\mathbf\cF\cR^{PD}_{g;c}$
 and indefinite in the space
 $\mathbf\cF\cR^{SR}_{g;c} $
).
The norm
\be
||b||=c\equiv \sqrt{a_{ij}(x)b^i(x)b^j(x)}
\ee
of the input 1-form  $b=b_iy^i$ is taken to be  an arbitrary
positive scalar subject to the restrictions: $0<c(x) <  1$ in the positive-definite space
 $\mathbf\cF\cR^{PD}_{g;c}$,
 and
$c(x) >  1$   in the indefinite space
 $\mathbf\cF\cR^{SR}_{g;c} $.
 In the latter space, we assume the 1-form $b$ to be time-like.

 \ses

If we  specify    the Riemannian metric tensor according to
\be
a_{ij} =  \fr1{c^2}b_ib_j  +  m u_{ij}
\ee
(with $ b^iu_{ij}=0$ and   $b^i=a^{ij}b_j$; $u_{ij}$ is the $(N-1)$-dimensional part
of Euclidean metric tensor, in compliance with (A.1)-(A.7))
and assume the scalars $c,m$ to be of the dependence
$
c=c(r),\,  m=m(r)$
on
the radius variable
$
r=\sqrt{u_{ij}x^ix^j}
$
(assuming also $c>0$ and $m\ne 0$),
we can readily conclude that
 the Riemannian covariant derivative
\be
\nabla_ib_j~=\D{b_j}{x^i}- b_na^n{}_{ij}
\ee
($a^n{}_{ij}$ stand for the Christoffel symbols constructed from the tensor $a_{ij}(x)$)
fulfills the equality
\be
\nabla_ib_j= \fr 1{c}(c_i b_j+c_jb_i)
\ee
(the respective calculations are presented in Appendix A);
$ c_i=\partial  c/\partial x^i$.
The right-hand part of (1.4)  does not involve  the function $m$.
It follows that
\be
y^iy^j\nabla_ib_j= \fr 2{c}b(yc),
\ee
where
$ (yc)=y^hc_h.$
We have
 \be
b^ic_i=0.
 \ee

{\it
Under the condition}
 (1.2)
 {\it  the Finsleroid-regular spaces cannot be of neither  Berwald case nor the Landsberg type.}
Obviously,  the Landsberg characteristic condition $\nabla_ib_j=kr_{ij}$ (derived and used in [2-5],
 and applied to cosmological consideration in [6]) is not realized when
(1.4) takes place.
The Berwald case would imply $\nabla_ib_j=0$ (see [1]), that is $c_i=0$.

At the same time, the
  Schwarzschild metric (which underlines the  gravitational field produced by static spherical-symmetric body)
belongs to the form (1.2).
Therefore, the pseudo-Finsleroid regular space
 $\mathbf\cF\cR^{SR}_{g;c} $
 taken together with the representations (1.2)-(1.4)
 provides us  with a straightforward continuation of the
Riemannian  Schwarzschild-metric framework into  the  pseudo-Finslerian domain
with respect to the pseudo-Finsleroid charge $g(x)$ (which thereby  plays the role of the characteristic  parameter
of extension).
The pseudo-Finsleroid axis  vector field $b_i(x)$ plays the role of the distribution of  the time variable.

Key objects involved are explicitly calculated below in Appendix A upon using the metric tensor (1.2).
Namely, calculating the associated Christoffel symbols
$  a^k{}_{ij}  $
leads to (A.31),
thereafter
the Riemannian curvature tensor
$
a_n{}^i{}_{km}
$
(see the definition (A.32)) is obtained according to (A.39),
and
the entailed Ricci tensor
$a_n{}^i{}_{im}$ is given by (A.40)-(A.43).
With these representations,
we examine the particular case:
\be
c=\fr{1+\fr {\xi}{4r}  } {1-\fr {\xi}{4r}}
\ee
(see (A.47))
and
\be
m=-\lf(1+\fr {\xi}{4r}\rg)^4
\ee
(see (A.49)),
obtaining
\be
a_n{}^i{}_{im}=0 \quad {\rm when} \quad N=4.
\ee
The respective
curvature tensor
$
a_n{}^i{}_{km}
$
is found to read (A.54).
The  vanishing (1.9) demonstrates that the metric tensor (1.2) with the choice (1.7) and (1.8)
fulfills the gravitational field equations in vacuum.
Raising forth the identification
\be
\xi=r_g,
\ee
where $r_g$ is the gravitational   radius of the massive static and spherical-symmetric body,
we may conclude that  in
the dimension $N=4$
the metric tensor (1.2) with the choice (1.7) and (1.8) is just
the Schwarzschild metric  tensor
(cf. [8,9]).

\ses

It should be noted that
$c<1$ of the positive-definite space
transforms to
$c>1$ in the relativistic counterpart
with the timelike vector $b_i$. The reason is that
the Riemannian metric
$$S^2=b^2+q^2$$
 of the positive-definite space must be replaced by
$$
S^2=b^2-q^2,
$$
 when treating the relativistic case.

In Appendix B, using the spray coefficients found in the previous work [1],
we evaluate the associated Finslerian curvature tensor $R^i{}_k$ under the only conditions that
$
\nabla_nb_m-\nabla_mb_n=0
$
 and
$   g=const,  $
simultaneously keeping the general case
$
c=c(x)
$
of the norm value $||b||$. The resultant representation is sufficiently simple and transparent
and, therefore,  opens  up the operative possibility to evaluate
the   tensor
\be
\rho_{ij}~:=\fr12(R_i{}^m{}_{mj}+R^m{}_{ijm})-\fr12g_{ij}R^{mn}{}_{nm}
\ee
and the current
\be
\rho^i~:=\rho^i{}_j y^j
\ee
(see [2,6])
which can be used to represent  the energy-momentum distribution in the respective pseudo-Finsleroid-regular
space-time.
Various relativistic and other applications are possible.


\ses\ses

\ses\ses

\ses\ses

\ses\ses

\setcounter{equation}{0}

\nin
{\large \bf Appendix A:  Spherical-symmetric Riemannian background  }

\ses

\ses

Let us start with the (pseudo)Euclidean  metric tensor $e_{ij}$, which inverse will be denoted by
 $e^{ij}$, so that   $e_{ij}  e^{jn}=\de_i{}^n$,
and introduce  the decomposition
\be
e_{ij}= e_i  e_j+\epsilon u_{ij}, \qquad \epsilon = \pm 1,
\ee
where $e_i$ is a unit vector and
\be
 \text{rank}(u_{ij})=N-1, \qquad y^iy^ju_{ij}\ge0.
\ee
The value $\epsilon=1$ corresponds to the positive-definite case; in the relativistic case we must take
$\epsilon=-1$.
If we construct
the contravariant vector
\be
e^i=e^{ij} e_j,
\ee
we obtain
\be
e_ie^i=1, \qquad
e^iu_{ij}=0.
 \ee
We get also
\be
u_i{}^j=u^{jm}u_{im},
 \qquad
u^{ij}=e^{im}e^{jn}u_{mn},
\ee
and
\be
   u^{ij}u_{jn}=\de^i{}_n-e^ie_n,
   \qquad
   u_i{}^j=\de_i{}^j-e_ie^j.
\ee

Let us set forth the identification
\be
b_i=e_i,
\ee
introduce two  scalars
$c>0$ and $m\ne0$,
and specify   the Riemannian metric tensor as follows:
\be
a_{ij}~: =  \fr1{c^2}b_ib_j  +  m u_{ij},
\ee
which inverse reads
\be
a^{ij}=\fr1{c^2}  b^ib^j  +  \fr1m u^{ij},
\ee
so that     $a_{ij}  a^{jn}=\de_i{}^n$.

{

We use the contravariant vector
\be
b^i=a^{ij}b_j,
\ee
\ses
which entails
 \be
b^iu_{ij}=0
 \ee
\ses
and
\be
b^ib_i=c^2,
\ee
\ses
together with
\be
b^i=c^2  e^i.
\ee

\ses

We introduce the radius variable
\be
r=\sqrt{u_{ij}x^ix^j},
\ee
\ses
obtaining
\be
n_i~:=\fr1{r}  u_{ij}x^j  = \D r{x^i},
\ee
\ses
\be
\D{n_i}{x^j}=\fr1{r}     \lf(u_{ij}-n_in_j\rg),
\ee
\ses
\be
n^i=u^{ij}n_j=\epsilon e^{ij}n_j,
\ee
and also
\be
n^in_i=1
\ee
together with
\be
n_ib^i=0.
\ee

We shall specify the scalars $c,m$ to be of the dependence
\be
c=c(r),  \qquad  m=m(r),
\ee
assuming $c>0$ and $m\ne 0$.
We shall also use the notation
\be
c_i=\D c{x^i} \equiv c'n_i
\ee
and
\ses
\be
c^i=a^{ij}c_j,
\ee
so that
\be
c^i=\fr1m c'n^i
\ee
and
 \be
b^ic_i=0.
 \ee

{

Under these conditions, we obtain
\be
\D{a_{ij}}{x^n}
=n_n\lf ( -\fr2{c^3}c'b_ib_j+ m' u_{ij}\rg),
\ee
from which it just follows that
$$
b_na^n{}_{ij}=\fr12 b^n
\Biggl[ n_i\lf(-\fr{2c'}{c^3} b_nb_j+ m' u_{nj} \rg)+    n_j\lf( -\fr2{c^3}c'b_ib_n+ m' u_{in}\rg)
+n_n\lf( \fr2{c^3}c'b_ib_j- m' u_{ij}\rg)  \Biggr],
$$
\ses
or
$$
b_na^n{}_{ij}=-   \fr 1{c}c'(n_i b_j+n_jb_i)
= -   \fr 1{c}(c_i b_j+c_jb_i).
$$
\ses
Thus for the Riemannian covariant derivative
\be
\nabla_ib_j~=\D{b_j}{x^i}- b_na^n{}_{ij}
\ee
we obtain the simple expression
\be
\nabla_ib_j= \fr 1{c}(c_i b_j+c_jb_i),
\ee
the right-hand part of which does not involve  the function $m$.
It follows that
\be
y^iy^j\nabla_ib_j= \fr 2{c}b(yc).
\ee
We use the notation
\be
 (yc)=y^hc_h.
\ee

      \ses

 With (A.25), we also obtain
$$
c_na^n{}_{ij}=\fr12 c^n
\Biggl[ n_i\lf(-\fr2{c^3}c' b_nb_j+ m' u_{nj} \rg)+    n_j\lf( -\fr2{c^3}c'b_ib_n+ m' u_{in}\rg)
+n_n\lf(\fr2{c^3}c'b_ib_j  -     m' u_{ij}\rg)  \Biggr]
$$
\ses
and derive
$$
c_na^n{}_{ij}=\fr12 \fr1m  c' \Biggl[ 2 m' n_in_j
+\fr2{c^3}c'b_ib_j  -  m' u_{ij}  \Biggr],
$$
\ses
so that
\be
\nabla_ic_j=c''n_i n_j
+c'\fr1{r}     \lf(u_{ij}-n_in_j\rg)
-\fr12 \fr1m  c' \Biggl[ 2 m' n_in_j
+\fr2{c^3}c'b_ib_j  -  m' u_{ij}  \Biggr].
\ee

{

Calculating the Christoffel symbols
$  a^k{}_{ij}  $
on the basis of (A.8) and (A.9) yields
straightforwardly that
$$
a^k{}_{ij}=   \fr12a^{kn}\Biggl[   n_i\lf ( -\fr2{c^3}c'b_nb_j+ m' u_{nj}\rg)
+  n_j\lf ( -\fr2{c^3}c'b_nb_i+ m' u_{ni}\rg)
+ n_n\lf (\fr2{c^3}c'b_ib_j- m' u_{ij}\rg)
\Biggr],
$$
\ses
or
\be
a^k{}_{ij}=  -\fr1{c^3}c'
b^k(  n_ib_j+n_jb_i)
+  \fr12 \fr1m \Biggl[
 m'n_iu_j{}^k
+    m'n_j u_i{}^k
+ n^k\lf ( \fr2{c^3}c'b_ib_j  -    m' u_{ij}\rg)
\Biggr].
\ee

\ses

We are aimed now to obtain
the Riemannian curvature tensor
\be
a_n{}^i{}_{km}=\D{a^i{}_{nm}}{x^k}-\D{a^i{}_{nk}}{x^m}+a^u{}_{nm}a^i{}_{uk}-a^u{}_{nk}a^i{}_{um}.
\ee

{

To this end we write (A.31) in the form
\be
a^i{}_{nm}=  -\fr1{c}c'
e^i(  n_nb_m+n_mb_n)
+  \fr12 \fr1m \Biggl[
 m'n_nu_m{}^i  +    m'n_m u_n{}^i
+ n^i\lf ( \fr2{c^3}c'b_nb_m -   m' u_{nm}\rg)
\Biggr]
\ee
and get
$$
\D{a^i{}_{nm}}{x^k}-\D{a^i{}_{nk}}{x^m}=
 -\lf(\fr1{c}c''  -  \fr1{c^2}c'c' \rg)
e^in_n( n_kb_m -n_mb_k)
$$

\ses

$$
 -\fr1{c}c'
 \fr1{r} \Bigl[    \lf(u_{nk}-  n_nn_k\rg) b_m -     \lf(u_{nm}-  n_nn_m\rg) b_k \Bigr]
 e^i
$$

\ses

$$
-  \fr12 \fr1{m^2}m' n_k \Biggl[
 m'n_nu_m{}^i    - n^i\lf ( -\fr2{c^3}c'b_nb_m+ m' u_{nm}\rg)
\Biggr]
$$

\ses

$$
+  \fr12 \fr1{m^2}m' n_m \Biggl[
 m'n_nu_k{}^i    - n^i\lf ( -\fr2{c^3}c'b_nb_k+ m' u_{nk}\rg)
\Biggr]
$$

\ses

$$
+  \fr12 \fr1m \Biggl[
 m''n_nn_k
+ m'     \fr1{r}   \lf(u_{nk}-  n_nn_k\rg) \Biggr]
 u_m{}^i
$$

\ses

$$
-  \fr12 \fr1m \Biggl[
 m''n_nn_m
+ m'     \fr1{r}   \lf(u_{nm}-  n_nn_m\rg) \Biggr]
 u_k{}^i
$$

         \ses
$$
+ \fr12 \fr1m \fr1r(u^i{}_k-n^in_k)\lf ( \fr2{c^3}c'b_nb_m -   m' u_{nm}\rg)
$$

         \ses
$$
- \fr12 \fr1m \fr1r(u^i{}_m-n^in_m)\lf ( \fr2{c^3}c'b_nb_k -   m' u_{nk}\rg)
$$

\ses

$$
+  \fr1m
 n^i\lf ( \fr1{c^3}c''-\fr3{c^4}c'c'\rg)
 b_n(n_kb_m -n_mb_k)
-  \fr12 \fr1m
 m''(n_k u_{nm} -n_m u_{nk})
n^i,
$$

{

\nin
or
$$
\D{a^i{}_{nm}}{x^k}-\D{a^i{}_{nk}}{x^m}=
- \fr{m'}m\fr1r
\lf( u_{mn}u_k{}^i -u_{kn}u_m{}^i\rg)
$$

\ses

\ses

$$
 -\fr1{c^2}
 \lf(\fr1{c}c''  -  \fr1{c^2}c'c' - \fr1r  \fr1{c}c'\rg)
n_n( n_kb_m -n_mb_k)
b^i
$$

\ses

$$
 -\fr1{c^3}c'
 \fr1{r} (  u_{nk} b_m -   u_{nm} b_k )
b^i
$$

\ses

$$
+  \fr12 \fr1{m} \fr{m'}m
\Biggl[
n_k   \lf ( -\fr2{c^3}c'b_nb_m+ m' u_{nm}\rg)
-  n_m \lf ( -\fr2{c^3}c'b_nb_k+ m' u_{nk}\rg)
\Biggr]
  n^i
$$

\ses

$$
+  \fr12 \fr1m \Biggl[
 m''   -   \fr1{r}  m'     -\fr{(m')^2}m  \Biggr]
n_n  \lf( n_k u_m{}^i -n_m u_k{}^i \rg)
$$

         \ses
$$
+ \fr1m \fr1r\fr1{c^3}c'b_n  \lf(b_mu^i{}_k  -b_ku^i{}_m  \rg)
$$

\ses

\ses

$$
+  \fr1m
\fr1{c^2}
\lf ( \fr1{c}c''-\fr3{c^2}c'c'  -  \fr1r \fr1c c'\rg)
 b_n(n_kb_m -n_mb_k)
 n^i
 $$

\ses

$$
-  \fr12 \fr1m
\lf( m'' -\fr1rm'\rg)(n_k u_{nm} -n_m u_{nk})
n^i.
$$

\ses

{

In this way we get
$$
\D{a^i{}_{nm}}{x^k}-\D{a^i{}_{nk}}{x^m}=
- \fr{m'}m\fr1r
\lf( u_{mn}u_k{}^i -u_{kn}u_m{}^i\rg)
$$

\ses

$$
 -\fr1{c^2}
 \lf(\fr1{c}c''  -  \fr1{c^2}c'c' - \fr1r  \fr1{c}c'\rg)
\Biggl[ n_n( n_kb_m -n_mb_k)  b^i  -   \fr1m b_n( n_kb_m -n_mb_k)  n^i  \Biggr]
$$

\ses

$$
 -\fr1{c^3}c'
 \fr1{r} (  u_{nk} b_m -   u_{nm} b_k )
b^i
-   \fr1{m} \fr{m'}m
\fr1{c^3}c'b_n(n_kb_m-n_mb_k)
  n^i
$$

\ses

$$
-  \fr12 \fr1m \Biggl[
 m''   -   \fr1{r}  m'     -\fr{(m')^2}m  \Biggr]
\Bigl(
n_n(n_mu_k{}^i -n_ku_m{}^i)
- (n_m u_{nk} -n_k u_{nm})     n^i
\Bigr)
$$

         \ses
\be
+ \fr1m \fr1r\fr1{c^3}c'b_n  \lf(b_mu^i{}_k  -b_ku^i{}_m  \rg)
-  \fr1m
\fr1{c^2}
\fr2{c^2}c'c'
 b_n(n_kb_m -n_mb_k)
 n^i.
\ee

\ses

{

Next,  we obtain the representation
$$
a^u{}_{nm}a^i{}_{uk}-a^u{}_{nk}a^i{}_{um}=
 -\fr1{c}c'
(  n_nb_m+n_mb_n)
\Biggl[
 -\fr1{c}c'     n_k  e^i
+  \fr1m
 \fr1{c^3}c     'b_k    n^i
\Biggr]
$$

\ses

$$
-\fr12 \fr1{c}c'\fr1m  m'
n_n
           n_mb_k
 e^i
$$

\ses

$$
 +\fr12 \fr{m'}m  n_mu_n{}^u
 \Biggl[
 -\fr{c'}{c}
e^i(  n_ub_k+n_kb_u)
+  \fr12 \fr1m \Biggl(
 m'n_uu_k{}^i  +    m'n_k u_u{}^i
+ n^i\lf ( \fr{2c'}{c^3}b_ub_k -   m' u_{uk}\rg)
\Biggr)
\Biggr]
$$

\ses

$$
+ \fr12 \fr1m
\lf ( \fr2{c^3}c'b_nb_m -   m' u_{nm}\rg)
 \Biggl[
 -\fr1{c}c'
e^i  b_k
+  \fr12 \fr1m
 m'u_k{}^i
\Biggr]
-[km]
$$

\ses

\ses

\ses

$$
=
\fr1{c^2}c'c'  n_n ( b_m   n_k -b_kn_m) e^i
 -\fr1{c^4}c'c'  \fr1m    b_n ( n_m b_k -n_kb_m )  n^i
 $$

\ses

$$
-\fr12 \fr1{c}c'\fr{m'}m
n_n
 (   n_mb_k-n_kb_m)
 e^i
$$

\ses

$$
-\fr12\fr{m'}m
\fr{c'}{c}
  n_n (n_mb_k -n_kb_m)    e^i
+ \fr14\lf(\fr{m'}m\rg)^2
\Bigl(
n_n(n_mu_k{}^i -n_ku_m{}^i)
- (n_m u_{nk} -n_k u_{nm})     n^i
\Bigr)
$$

\ses

$$
+\fr12 \fr1m
\fr{c'}{c^3}   \fr{m'}m    b_n(b_m    u_k{}^i    -b_k    u_m{}^i  )
-\fr12\fr{m'}m
 u_{nm}   \Biggl[   -\fr{c'}{c}   e^i  b_k +  \fr12\fr{m'}m u_k{}^i  \Biggr]
+\fr12\fr{m'}m
 u_{nk}   \Biggl[   -\fr{c'}{c}   e^i  b_m +  \fr12\fr{m'}m u_m{}^i  \Biggr].
$$

\ses

{

Eventually,
$$
a^u{}_{nm}a^i{}_{uk}-a^u{}_{nk}a^i{}_{um}=
$$

\ses

$$
=
\fr1{c^4}c'c'
 \Biggl[
  n_n    (   b_mn_k-b_kn_m)b^i
-  b_n ( n_m b_k -n_kb_m ) \fr1m    n^i
\Biggr]
$$

\ses

\ses

$$
- \fr1{c^3}c'\fr{m'}m
n_n
 (   n_mb_k-n_kb_m)
b^i
$$

\ses

\ses

$$
+ \fr14\lf(\fr{m'}m\rg)^2
\Bigl(
n_n(n_mu_k{}^i -n_ku_m{}^i)
- (n_m u_{nk} -n_k u_{nm})     n^i
\Bigr)
$$

\ses

\ses
$$
+\fr12 \fr1m
\fr{c'}{c^3}   \fr{m'}m    b_n(b_m    u_k{}^i    -b_k    u_m{}^i  )
$$

\ses

\ses

\be
-\fr12\fr{m'}m
 u_{nm}   \Biggl[   -\fr1{c^3}c'   b^i  b_k +  \fr12\fr{m'}m u_k{}^i  \Biggr]
+\fr12\fr{m'}m
 u_{nk}   \Biggl[   -\fr1{c^3}c'   b^i  b_m +  \fr12\fr{m'}m u_m{}^i  \Biggr].
\ee

{


Adding (A.35) to (A.34) we find the curvature tensor (A.32):
$$
a_n{}^i{}_{km}=
- \fr{m'}m
\lf(\fr14\fr{m'}m
+\fr1r\rg)
\lf( u_{mn}u_k{}^i -u_{kn}u_m{}^i\rg)
$$

\ses

\ses

$$
 -\fr1{c^2}
 \lf(\fr1{c}c''  -  \fr2{c^2}c'c' - \fr1r  \fr1{c}c'    -\fr{c'}c\fr{m'}m  \rg)
\Biggl[ n_n( n_kb_m -n_mb_k)  b^i  -   \fr1m b_n( n_kb_m -n_mb_k)  n^i  \Biggr]
$$

\ses

\ses

$$
-  \fr12 \fr1m \Biggl[
 m''   -   \fr1{r}  m'     -\fr32  \fr{(m')^2}m  \Biggr]
\Bigl(
n_n(n_mu_k{}^i -n_ku_m{}^i)
- (n_m u_{nk} -n_k u_{nm})     n^i
\Bigr)
$$

\ses

\ses

\be
+\fr12
\fr{c'}{c^3} \lf(  \fr{m'}m   +\fr2r\rg)
\Biggl[ \fr1m  b_n(b_m    u_k{}^i    -b_k    u_m{}^i  )  -(b_mu_{nk}-b_ku_{nm})b^i\Biggr].
\ee

\ses

Inserting here
\be
mu_{mn}=a_{mn}-\fr1{c^2}b_mb_n
\ee
and
\be
u_i{}^k=\de_i{}^k -\fr1{c^2}b_ib^k
\ee

{

\nin
yields   the result
$$
a_n{}^i{}_{km}=
-
\fr1m   \fr{m'}m
\lf(\fr14\fr{m'}m
+\fr1r\rg)
\lf( a_{mn}\de_k{}^i -a_{kn}\de_m{}^i\rg)
$$

\ses

\ses

$$
+\fr1{c^2} \fr{m'}m
\lf(\fr14\fr{m'}m
+\fr1r\rg)
\lf(\fr1m( a_{mn}b_kb^i -a_{kn}b_mb^i)    +  b_mb_n\de_k{}^i -b_kb_n\de_m{}^i\rg)
$$

\ses

\ses

$$
 -\fr1{c^2}
 \lf(\fr1{c}c''  -  \fr2{c^2}c'c' - \fr1r  \fr1{c}c'     -\fr{c'}c\fr{m'}m  \rg)
\Biggl[ n_n( n_kb_m -n_mb_k)  b^i  -   \fr1m b_n( n_kb_m -n_mb_k)  n^i  \Biggr]
$$

\ses

\ses

$$
-  \fr12 \fr1m \Biggl[
 m''   -   \fr1{r}  m'     -\fr32  \fr{(m')^2}m  \Biggr]
\Bigl(
n_n(n_mu_k{}^i -n_ku_m{}^i)
- (n_m u_{nk} -n_k u_{nm})     n^i
\Bigr)
$$

\ses

\ses

\be
+\fr12
\fr{c'}{c^3} \lf(  \fr{m'}m  +\fr2r\rg)
\Biggl[ \fr1m  b_n(b_m    u_k{}^i    -b_k    u_m{}^i  )  -(b_mu_{nk}-b_ku_{nm})b^i\Biggr].
\ee

{

Contracting over two indices leads to the representation
$$
a_n{}^i{}_{im}=
- \fr{m'}m
\lf(\fr14\fr{m'}m  +\fr1r\rg)   (N-2) u_{nm}
$$

\ses

\ses

$$
+\fr1{c^2}
 \lf(\fr1{c}c''  -  \fr2{c^2}c'c' - \fr1r  \fr1{c}c'    -\fr{c'}c\fr{m'}m  \rg)
\lf[ c^2n_nn_m  +   \fr1m b_nb_m \rg]
$$

\ses

\ses

$$
-  \fr12 \fr1m \lf(
 m''   -   \fr1{r}  m'     -\fr32  \fr{(m')^2}m  \rg)
\lf[  u_{nm}+(N-3)n_nn_m
\rg]
$$

\ses

\ses

$$
 +\fr12 \fr{c'}{c^3} \lf(  \fr{m'}m   +\fr2r\rg)
\lf[c^2u_{nm}+(N-1) \fr1m  b_nb_m\rg],
$$

{

\nin
which can be written as
\be
a_n{}^i{}_{im}=n_1u_{nm}
+
\fr1{c^2}\fr1mn_2b_mb_n
+n_3n_mn_n
\ee
with
\be
n_1=-(N-2) \fr{m'}m
\lf(\fr14\fr{m'}m  +\fr1r\rg)
-  \fr12 \fr1m \lf(
 m''   -   \fr1{r}  m'     -\fr32  \fr{(m')^2}m  \rg)
+\fr12
\fr{c'}{c} \lf(  \fr{m'}m   +\fr2r\rg),
\ee

\ses

\be
n_2=
 \lf(\fr1{c}c''  -  \fr2{c^2}c'c' - \fr1r  \fr1{c}c'  -\fr{c'}c\fr{m'}m  \rg)
 +\fr12
\fr{c'}{c} \lf(  \fr{m'}m   +\fr2r\rg)
(N-1),
 \ee
\ses
and
\be
n_3=
 \lf(\fr1{c}c''  -  \fr2{c^2}c'c' - \fr1r  \fr1{c}c'  -\fr{c'}c\fr{m'}m  \rg)
-  \fr12 \fr1m \lf(
 m''   -   \fr1{r}  m'     -\fr32  \fr{(m')^2}m  \rg)
(N-3).
 \ee

{

Let us make the substitution
\be
c=y\lf(\fr {\xi}{4r}\rg), \qquad
m=z\lf(\fr {\xi}{4r}\rg),
\qquad \xi = const,
\ee
so that
\be
c'=-\fr {\xi}{4r^2}y', \qquad   m'=-\fr {\xi}{4r^2}z',
\ee
\ses
and
\be
c''= \fr {\xi}{2r^3}y'    +\fr{ {\xi}^2}{16r^4}y'', \qquad    m''= \fr {\xi}{2r^3}z'    +\fr{ {\xi}^2}{16r^4}z'',
\ee
\ses
$$
\fr1{c}c''  -  \fr2{c^2}c'c' - \fr1r  \fr1{c}c'=
\fr {\xi}{2r^3y}y'    +\fr{ {\xi}^2}{16r^4y}y''
- 2\fr {\xi}{4r^2y}y'\fr {\xi}{4r^2y}y'
+   \fr {\xi}{4r^3y}y',
$$

{

\nin
and examine the particular case
\be
c=\fr{1+\fr {\xi}{4r}  } {1-\fr {\xi}{4r}}, \qquad {\xi}>0.
\ee
\ses
We have
\be
y'=  2\fr1   {\lf(1-\fr {\xi}{4r}\rg)^2},  \qquad
y''=  4\fr1   {\lf(1-\fr {\xi}{4r}\rg)^3},
\ee
\ses
whence
$$
\fr1{c}c''  -  \fr2{c^2}c'c' - \fr1r  \fr1{c}c'=
3\fr {\xi}{4r^3y}y'   +
\fr{ {\xi}^2}{16r^4y}y''
- 2\fr {\xi}{4r^2y}y'\fr {\xi}{4r^2y}y'
$$

\ses

\ses

$$
=
6\fr 1{r^2}  \fr{\fr {\xi}{4r}}
   {\lf(1+\fr {\xi}{4r}\rg)\lf(1-\fr {\xi}{4r}\rg)}
+
\fr4{r^2}
\fr{\lf(\fr {\xi}{4r}\rg)^2  }
{\lf(1+\fr {\xi}{4r}\rg)\lf(1-\fr {\xi}{4r}\rg)^2}
-
\fr8{r^2}
\fr{\lf(\fr {\xi}{4r}\rg)^2 }
{\lf(1+\fr {\xi}{4r}\rg)^2\lf(1-\fr {\xi}{4r}\rg)^2}
$$

\ses

\ses

$$
=
6\fr 1{r^2}  \fr{\fr {\xi}{4r}}
   {\lf(1+\fr {\xi}{4r}\rg)\lf(1-\fr {\xi}{4r}\rg)}
-
\fr4{r^2}
\fr{\lf(\fr {\xi}{4r}\rg)^2  }
{\lf(1+\fr {\xi}{4r}\rg)^2\lf(1-\fr {\xi}{4r}\rg)}
$$

\ses

$$
=- \fr 1{r^2}
\fr{\fr {\xi}{4r}}
   {\lf(1+\fr {\xi}{4r}\rg)\lf(1-\fr {\xi}{4r}\rg)}
   \Biggl[-6+ 4 \fr{\fr {\xi}{4r}}
   {\lf(1+\fr {\xi}{4r}\rg)}
   \Biggr].
   $$

{

We shall confine the treatment to  the relativistic $\epsilon=-1$:
\be
m=-\lf(1+\fr {\xi}{4r}\rg)^4,
\ee
\ses
obtaining
$$
\fr{m'}m+\fr2r
=
-4\fr 1{r}  \fr{\fr {\xi}{4r}}
   {\lf(1+\fr {\xi}{4r}\rg)}
+\fr2r
   $$
\ses
and
$$
\fr12\fr{c'}c
\lf(\fr{m'}m+\fr2r\rg)
=
-
\fr1r
\fr{\fr {\xi}{4r}}
   {\lf(1+\fr {\xi}{4r}\rg)\lf(1-\fr {\xi}{4r}\rg)}
   \Biggl[-\fr 4{r}  \fr{\fr {\xi}{4r}}      {\lf(1+\fr {\xi}{4r}\rg)}   +\fr2r   \Biggr]
=
-\fr 2{r^2}  \fr{\fr {\xi}{4r}}      {\lf(1+\fr {\xi}{4r}\rg)^2},
   $$

\ses

\ses

$$
\fr{m'}m
\lf(\fr14\fr{m'}m  +\fr1r\rg)
=
-\fr 4{r}  \fr{\fr {\xi}{4r}}      {\lf(1+\fr {\xi}{4r}\rg)}
   \Biggl[-\fr 1{r}  \fr{\fr {\xi}{4r}}      {\lf(1+\fr {\xi}{4r}\rg)}   +\fr1r   \Biggr]
=
-
\fr 4{r^2}  \fr{\fr {\xi}{4r}}      {\lf(1+\fr {\xi}{4r}\rg)^2},
$$
\ses
and also
$$
\lf(\fr{m'}m\rg)'=-\fr2r  \fr{m'}m
+\fr {\xi}{4r^2}\fr{m'}m
\fr{1}      {\lf(1+\fr {\xi}{4r}\rg)}.
$$
\ses
Therefore,
$$
\fr1m
\lf( m''   -   \fr1{r}  m'     -\fr32  \fr{(m')^2}m\rg)
=
\lf(\fr{m'}m\rg)'-\fr1r \fr{m'}m
- \fr12 \fr{(m')^2}{m^2}
$$

\ses

$$
= \fr{m'}m
\Biggl[
-\fr2r
+\fr {\xi}{4r^2}
\fr{1}      {\lf(1+\fr {\xi}{4r}\rg)}
-
\fr1r
+2
\fr {\xi}{4r^2}
\fr{1}      {\lf(1+\fr {\xi}{4r}\rg)}
\Biggr]
$$

\ses

$$
= \fr{m'}m
\fr3r\Biggl[
-1
+\fr {\xi}{4r}
\fr{1}      {\lf(1+\fr {\xi}{4r}\rg)}
\Biggr]
= -\fr{m'}m   \fr3r
\fr1      {\lf(1+\fr {\xi}{4r}\rg)}
=\fr{12}{r^2}
\fr {\lf(\fr {\xi}{4r}\rg)}
      {\lf(1+\fr {\xi}{4r}\rg)^2},
$$
\ses
whence
we have
\be
n_1=0.
\ee

{

Thereafter, we obtain
$$
 \fr1{c}c''  -  \fr2{c^2}c'c' - \fr1r  \fr1{c}c'-\fr{c'}c\fr{m'}m
 =
 $$

\ses

$$
- \fr 1{r^2}
\fr{\fr {\xi}{4r}}
   {\lf(1+\fr {\xi}{4r}\rg)\lf(1-\fr {\xi}{4r}\rg)}
   \Biggl[-6+ 4 \fr{\fr {\xi}{4r}}
   {\lf(1+\fr {\xi}{4r}\rg)}
   \Biggr]
   -
\fr2r
\fr{\fr {\xi}{4r}}
   {\lf(1+\fr {\xi}{4r}\rg)\lf(1-\fr {\xi}{4r}\rg)}
\fr 4{r}  \fr{\fr {\xi}{4r}}      {\lf(1+\fr {\xi}{4r}\rg)}
  $$

\ses

\ses

$$
=6 \fr 1{r^2}
\fr{\fr {\xi}{4r}}
   {\lf(1+\fr {\xi}{4r}\rg)\lf(1-\fr {\xi}{4r}\rg)}
   \Biggl[1-2 \fr{\fr {\xi}{4r}}
   {\lf(1+\fr {\xi}{4r}\rg)}
   \Biggr]
=
6 \fr 1{r^2}
\fr{\fr {\xi}{4r}}
   {\lf(1+\fr {\xi}{4r}\rg)^2},
   $$
whence
\be
n_2=n_3=0.
\ee
\ses

{

The conclusions (A.50) and (A.51) tell us that
\be
a_n{}^i{}_{im}=0 \quad {\rm when} \quad N=4.
\ee

Under these conditions, the tensor (A.39) reduces to
$$
a_n{}^i{}_{km}=
 \fr 2{r^2}
\fr{\fr {\xi}{4r}}
   {\lf(1+\fr {\xi}{4r}\rg)^2}
\Biggl[
2
\lf( u_{mn}u_k{}^i -u_{kn}u_m{}^i\rg)
$$

\ses

\ses

$$
 -\fr3{c^2}
\Biggl( n_n( n_kb_m -n_mb_k)  b^i  -   \fr1m b_n( n_kb_m -n_mb_k)  n^i  \Biggr)
$$

\ses

\ses

$$
- 3
\Bigl(
n_n(n_mu_k{}^i -n_ku_m{}^i)
- (n_m u_{nk} -n_k u_{nm})     n^i
\Bigr)
$$

\ses

\ses

\be
-\fr1{c^2}
\Biggl( \fr1m  b_n(b_m    u_k{}^i    -b_k    u_m{}^i  )  -(b_mu_{nk}-b_ku_{nm})b^i\Biggr)
\Biggr],
\ee

{

\nin
which can also be written as
$$
a_n{}^i{}_{km}=
 \fr 2{r^2}
\fr{\fr {\xi}{4r}}
   {\lf(1+\fr {\xi}{4r}\rg)^2}
\Biggl[
2
\lf( u_{mn}u_k{}^i -u_{kn}u_m{}^i\rg)
$$

\ses

\ses

$$
- 3
\Bigl(
n_n(n_mu_k{}^i -n_ku_m{}^i)
- (n_m u_{nk} -n_k u_{nm})     n^i
\Bigr)
$$

\ses

\ses

\be
-\fr1{c^2}
\Biggl( \fr1m  b_n\Bigl(b_m  (  u_k{}^i -3n_kn^i)   -b_k   ( u_m{}^i-3n_mn^i)  \Bigr)
 -\Bigl(b_m(u_{nk}-3n_nn_k)     -b_k(u_{nm}-3n_nn_m)\Bigr)b^i\Biggr)
\Biggr].
\ee

{

Contracting the last tensor by the vector $b$ yields the simple result, namely,
\be
b^ma_n{}^i{}_{km}=
 \fr 2{r^2}
\fr{\fr {\xi}{4r}}
   {\lf(1+\fr {\xi}{4r}\rg)^2}
\Biggl[
-
\lf( \fr1m  b_n   ( u_k{}^i-3n_kn^i)     -(u_{nk}-3n_nn_k)b^i\rg)
\Biggr]
\ee
\ses
and
\be
b^na_n{}^i{}_{km}=
 \fr 2{r^2}
\fr{\fr {\xi}{4r}}
   {\lf(1+\fr {\xi}{4r}\rg)^2}
\Biggl[
-\fr1m\Bigl(b_m(u_k{}^i-3n_kn^i)-b_k(u_m{}^i-3n_mn^i)  \Bigr)
\Biggr],
\ee
\ses
from which it follows that
$$
(bb^mb^n-b^my^n-b^ny^m)
a_n{}^i{}_{km}=
$$

\ses

\be
 \fr 2{r^2}
\fr{\fr {\xi}{4r}}
   {\lf(1+\fr {\xi}{4r}\rg)^2}
\Biggl[
- (u_{nk}-3n_nn_k)b^iy^n
+\fr1mb(u_k{}^i-3n_kn^i)   -\fr1m b_k(u_m{}^i-3n_mn^i)y^m
\Biggr].
\ee

{

\ses

\ses

\ses\ses

\setcounter{equation}{0}

\nin
{\large \bf Appendix B:  Curvature tensor of the  space  ${\mathbf\cF\cR^{PD}_{g;c} } $  }

\ses\ses

Below the evaluations are restricted by the particular conditions
\be
\nabla_nb_m-\nabla_mb_n=0, \quad g=const,
\ee
simultaneously keeping the general case
\be
c=c(x)
\ee
of the norm value $||b||$.

In terms of the convenient notation
\be
(ys)=y^jy^h\nabla_jb_h
\ee
the spray coefficients
 of the  space  ${\mathbf\cF\cR^{PD}_{g;c} } $
 read
\be
G^i=
\fr  g{\nu}
(ys)
v^i
+a^i{}_{km}y^my^k
\ee
(see [1]),
\ses
entailing
\be
G^i{}_k=
-  \fr1{\nu}
\nu_k
\fr  g{\nu}
(ys)
v^i
+        2 \fr  g{\nu}   s_k
v^i
+
\fr  g{\nu}
(ys)
r^i{}_k
+2a^i{}_{km}y^m,
\ee
\ses
where $s_k=y^h\nabla_kb_h$ and
\be
\nu_k=\fr1qv_k+(1-c^2)gb_k.
\ee
The derivative
\be
\D{\lf(\fr1{\nu}\nu_k\rg)}   {y^m}=-\fr1{\nu^2}\nu_k\nu_m+\fr1{\nu}\fr1q\eta_{km}
\ee
will be applied.

{

Differentiating (B.4) with respect to $y^m$ yields the result
$$
G^i{}_{km}=
-  \fr1{\nu}
\nu_k
\Biggl(
-  \fr1{\nu}
\nu_m
\fr  g{\nu}
(ys)
v^i
+        2 \fr  g{\nu}   s_m
v^i
+
\fr  g{\nu}
(ys)
r^i{}_m
\Biggr)
-
\Biggl(
-\fr1{\nu^2}\nu_k\nu_m+\fr1{\nu}\fr1q\eta_{km}
\Biggr)
\fr  g{\nu}
(ys)
v^i
$$

\ses

$$
-   2 \fr  g{\nu^2} \nu_m  s_k   v^i
+   2 \fr  g{\nu}   (\nabla_m b_k)   v^i
+   2 \fr  g{\nu}   s_k   r^i{}_m
$$

\ses

$$
- \fr  g{\nu^2}  (ys)\nu_m  r^i{}_k
+ 2\fr  g{\nu}  s_m  r^i{}_k
+\De,
$$

{

\nin
which can conveniently be written as follows:
$$
G^i{}_{km}=
-
\Biggl(
-\fr2{\nu^2}\nu_k\nu_m+\fr1{\nu}\fr1q\eta_{km}
\Biggr)
\fr  g{\nu}
(ys)
v^i
+   2 \fr  g{\nu}   (\nabla_m b_k)   v^i
$$

\ses

\be
-   2 \fr  g{\nu^2} (\nu_m  s_k +\nu_ks_m)  v^i
+   2 \fr  g{\nu}  ( s_k   r^i{}_m  +   s_m   r^i{}_k)
- \fr  g{\nu^2}  (ys)(\nu_m  r^i{}_k+\nu_k  r^i{}_m)
+2a^i{}_{km}.
\ee

{

Next, we perform required differentiation with respect to $x^k$, obtaining
$$
2\D{\bar G^i}{x^k}=\D{ G^i}{x^k}
=
-\fr{g}{\nu^2}(ys)
\Bigl(q_{,k}+g(1-c^2)b_{j,k}y^j  - 2gbcc_k\Bigr)v^i
$$

\ses

$$
+
\fr{g}{\nu}
v^iy^my^n \nabla_k\nabla_mb_n
-\fr{g}{\nu}(ys)
(b_{j,k}y^jb^i+bb^i_{,k})
+\De.
$$
\ses
Since
\be
q_{,k}=-\fr1qbb_{j,k}y^j  =-\fr1qb s_k+\De,
\ee
\ses
we can write
$$
\D{ G^i}{x^k}
=
\fr{g}{q\nu^2}(ys)
\Bigl(b-gq(1-c^2)\Bigr)s_kv^i
+
\fr{g}{\nu^2}(ys)
2gbcc_kv^i
$$

\ses

\be
+
\fr{g}{\nu}
v^iy^my^n \nabla_k\nabla_mb_n
-\fr{g}{\nu}(ys)
(s_kb^i+b\nabla_kb^i)
+\De.
\ee

\ses

{

Also,
$$
y^j\D{ G^i}{x^j}=
\fr{g}{q\nu^2}(ys)
\Bigl(b-gq(1-c^2)\Bigr)(ys)v^i
+
\fr{g}{\nu^2}(ys)
2gbc(yc)v^i
$$

\ses

$$
+
\fr{g}{\nu}
v^iy^jy^my^n \nabla_j\nabla_mb_n
-\fr{g}{\nu}(ys)
((ys)b^i+bs^i)
+\De
$$
\ses
with
the vector
\be
e_k=\fr b{q^2}v_k-b_k,
\ee
\ses
which obeys the equality
$$
\D{e_k}{y^j}=\fr b{q^2}\eta_{kj}-\fr {1}{q^2}v_ke_j.
$$
\ses

{

\ses

After that, we derive
the equality
$$
\nu^2
\D{\Bigl(y^j\D{ G^i}{x^j}\Bigr)}{y^k}
=
-2\fr{g}{q\nu}  \nu_k
(ys)
\Bigl(b-gq(1-c^2)\Bigr)(ys)v^i
-\fr{g}{q^3}v_k(ys)
\Bigl(b-gq(1-c^2)\Bigr)(ys)v^i
$$

\ses

$$
+2\fr{g}{q}(ys)
\Bigl(b-gq(1-c^2)\Bigr)s_kv^i
+
\fr{g}{q}(ys)
\Bigl(b-gq(1-c^2)\Bigr)(ys)r^i{}_k
$$

\ses

$$
+\fr{g}{q}(ys)
\Bigl(b_k-g\fr1qv_k(1-c^2)\Bigr)(ys)v^i
$$

\ses                 \ses

$$
-4 \fr1{\nu} \nu_k
g(ys)
gbc(yc)v^i
+
g(ys)
2gb_kc(yc)v^i
+
g(ys)
2gbcc_kv^i
$$

\ses

$$
+
g(ys)
2gbc(yc)r^i{}_k
+
gs_k
2gbc(yc)v^i
$$

\ses

$$
-
      {g}{\nu_k}
v^iy^jy^my^n \nabla_j\nabla_mb_n
+
{g}{\nu}
y^jy^my^n \nabla_j\nabla_mb_nr^i{}_k
$$

\ses

$$
+
{g}{\nu}
v^i\Bigl[\de^j{}_ky^my^n + 2y^jy^m\de^n{}_k\Bigr]
 \nabla_j\nabla_mb_n
$$

             \ses

                       \ses

$$
+{g}{\nu}_k(ys)
((ys)b^i+bs^i)
-{g}{\nu}s_k
((ys)b^i+bs^i)
-{g}{\nu}(ys)
(s_kb^i+b_ks^i)
+\De.
$$

{

With these observations, we find
$$
\nu^2
\Biggl[
3\D{ G^i}{x^k}
-
\D{\Bigl(y^j\D{ G^i}{x^j}\Bigr)}{y^k}
\Biggr]
=
$$

\ses          \ses

$$
3\fr{g}{q}(ys)
\Bigl(b-gq(1-c^2)\Bigr)s_kv^i
+
3{g}(ys)
2gbcc_kv^i
$$

\ses

$$
+
3{g}{\nu}
v^iy^my^n \nabla_k\nabla_mb_n
-3{g}{\nu}(ys)
s_kb^i
-3{g}{\nu}(ys)
b\nabla_kb^i
$$


\ses

\ses

\ses

$$
+2\fr{g}{q\nu}  \nu_k
(ys)
\Bigl(b-gq(1-c^2)\Bigr)(ys)v^i
+\fr{g}{q^3}v_k(ys)
\Bigl(b-gq(1-c^2)\Bigr)(ys)v^i
$$

\ses

$$
-2\fr{g}{q}(ys)
\Bigl(b-gq(1-c^2)\Bigr)s_kv^i
-
\fr{g}{q}(ys)
\Bigl(b-gq(1-c^2)\Bigr)(ys)r^i{}_k
$$

\ses

$$
-\fr{g}{q}(ys)
\Bigl(b_k-g\fr1qv_k(1-c^2)\Bigr)(ys)v^i
$$

\ses                 \ses

$$
+4 \fr1{\nu} \nu_k
g(ys)
gbc(yc)v^i
-
g(ys)
2gb_kc(yc)v^i
-
g(ys)
2gbcc_kv^i
$$

\ses

$$
-
g(ys)
2gbc(yc)r^i{}_k
-
gs_k
2gbc(yc)v^i
$$

\ses

$$
+
      {g}{\nu_k}
v^iy^jy^my^n \nabla_j\nabla_mb_n
-
{g}{\nu}
y^jy^my^n \nabla_j\nabla_mb_nr^i{}_k
$$

\ses

$$
-
{g}{\nu}
v^i\Bigl[\de^j{}_ky^my^n + 2y^jy^m\de^n{}_k\Bigr]
 \nabla_j\nabla_mb_n
$$

             \ses

                       \ses

$$
-{g}{\nu}_k(ys)
((ys)b^i+bs^i)
+{g}{\nu}s_k
((ys)b^i+bs^i)
+{g}{\nu}(ys)
(s_kb^i+b_ks^i)
+\De
$$

{

\ses          \ses

$$
=
2\fr{g}{q}(ys)
\Bigl(b-gq(1-c^2)\Bigr)s_kv^i
+
4{g}(ys)
gbcc_kv^i
$$

\ses

$$
+
2{g}{\nu}
v^iy^my^n \nabla_k\nabla_mb_n
-{g}{\nu}(ys)
s_kb^i
-3{g}{\nu}(ys)
b\nabla_kb^i
$$


\ses

            \ses

$$
+2\fr{g}{q\nu}  \nu_k
(ys)
\Bigl(b-gq(1-c^2)\Bigr)(ys)v^i
+\fr{g}{q^3}v_k(ys)
\Bigl(b-gq(1-c^2)\Bigr)(ys)v^i
$$

\ses

$$
-
\fr{g}{q}(ys)
\Bigl(b-gq(1-c^2)\Bigr)(ys)r^i{}_k
$$

\ses

$$
-\fr{g}{q}(ys)
\Bigl(b_k-g\fr1qv_k(1-c^2)\Bigr)(ys)v^i
$$

\ses                 \ses

$$
+
\Biggr(4 \fr1{\nu} \nu_k
g(ys)
gbcv^i
-
2g(ys)
gb_kcv^i
-
2g(ys)
gbcr^i{}_k
-
2gs_k
gbcv^i
\Biggr)
(yc)
$$

\ses

$$
-
g
\Bigl(y^jy^my^n \nabla_j\nabla_mb_n\Bigr)
(\nu r^i{}_k -{\nu_k}  v^i  )
-
2{g}{\nu}
\Bigl(y^jy^m   \nabla_j\nabla_mb_k \Bigr)
v^i
$$

             \ses

                       \ses

$$
-g{\nu}_k(ys)   (ys)b^i
+
\Bigr[
-gb {\nu}_k(ys)
+{g}{\nu}s_k
b
+ g{\nu}(ys)
b_k
\Bigr]
s^i
+\De.
$$

{

Finally,
$$
\nu^2
\lf[
3\D{ G^i}{x^k}
-
\D{\Bigl(y^j\D{ G^i}{x^j}\Bigr)}{y^k}
\rg]
=
$$

\ses

\ses

$$
-
g
\Bigl(y^jy^my^n \nabla_j\nabla_mb_n\Bigr)
(\nu r^i{}_k -{\nu_k}  v^i  )
+
2{g}{\nu}
y^jy^m \Bigl(  \nabla_k\nabla_mb_j -  \nabla_j\nabla_mb_k \Bigr)
v^i
$$

             \ses

                       \ses

$$
-3g{\nu}(ys)  b\nabla_kb^i
+
g  \Bigr[
{\nu}s_k-{\nu}_k(ys)
+
\fr1b {\nu}(ys)  b_k
\Bigr]
\Bigl(b s^i- (ys)b^i\Bigr)
+  \fr1b
g {\nu}(ys) (ys) b_k   b^i
$$

       \ses
                \ses

$$
+
\Biggr(4 \fr1{\nu} \nu_k
g(ys)
gbcv^i
-
2g(ys)
gb_kcv^i
-
2g(ys)
gbcr^i{}_k
-
2gs_k
gbcv^i
\Biggr)
(yc)
+
4{g}(ys)
gbc c_k    v^i
$$

\ses

\ses

$$
       +2\fr{g}{q}(ys)
\Bigl(b-gq(1-c^2)\Bigr)s_kv^i
$$

\ses

            \ses

$$
+2\fr{g}{q\nu}
(ys)
\Bigl(b-gq(1-c^2)\Bigr)(ys)  \nu_k   v^i
$$

\ses

\ses

$$
-
\fr{g}{q}(ys)
\Bigl(b-gq(1-c^2)\Bigr)(ys)\eta^i{}_k
$$

\ses

\ses

\be
-\fr{g}{q}(ys)
\Bigl(b_k-g\fr1qv_k(1-c^2)\Bigr)(ys)v^i
+\De.
\ee

{

After that, we evaluate
$$
\nu^4
G^i{}_jG^j{}_k=
\Bigl[
-
\nu_j
 g
(ys)
v^i
+        2  g{\nu}   s_j
v^i
+
  g{\nu}
(ys)
r^i{}_j
\Bigr]
\Bigl[
-
\nu_k
  g
(ys)
v^j
+        2   g{\nu}   s_k
v^j
+
  g{\nu}
(ys)
r^j{}_k
\Bigr]
=
$$

    \ses

$$
-
\nu_j
 g
(ys)
v^i
\Bigl[
-
\nu_k
  g
(ys)
v^j
+        2   g{\nu}   s_k
v^j
+
  g{\nu}
(ys)
r^j{}_k
\Bigr]
$$

\ses

    \ses

$$
+        2  g{\nu}   s_j
v^i
\Bigl[
-
\nu_k
  g
(ys)
v^j
+        2   g{\nu}   s_k
v^j
+
  g{\nu}
(ys)
r^j{}_k
\Bigr]
$$

\ses

    \ses

$$
+
  g{\nu}
(ys)
r^i{}_j
\Bigl[
-
\nu_k
  g
(ys)
v^j
+        2   g{\nu}   s_k
v^j
+
  g{\nu}
(ys)
r^j{}_k
\Bigr]
$$

{

\ses
\nin
and arrive at
$$
2\nu^4G^mG^i{}_{km}=
2  g
(ys)
v^m\Biggl(
2\nu_k\nu_m-{\nu}\fr1q\eta_{km}
\Biggr)
  g
(ys)
v^i
+  2  g
(ys)
v^m 2  g{\nu}^2   (\nabla_m b_k)   v^i
$$

\ses

$$
- 2  g{\nu}
(ys)
v^m
      2  g (\nu_m  s_k +\nu_ks_m)  v^i
+ 2  g{\nu}^2
(ys)
v^m         2  g ( s_k   r^i{}_m  +   s_m   r^i{}_k)
$$

\ses

\be
-   g{\nu}
(ys)
v^m    g  (ys)(\nu_m  r^i{}_k+\nu_k  r^i{}_m)
+\De.
\ee

{

Therefore,
$$
2\nu^4G^jG^i{}_{kj}
-
\nu^4
G^i{}_jG^j{}_k
=
$$

\ses

\ses

$$
2  g^2
(ys)
v^m\Biggl(
2\nu_k\nu_m-{\nu}\fr1q\eta_{km}
\Biggr)
(ys)
v^i
+  4  g^2
(ys)
v^m   {\nu}^2   (\nabla_m b_k)   v^i
$$

\ses

$$
- 4  g^2   \nu
(ys)
v^m
   (\nu_m  s_k +\nu_ks_m)  v^i
+ 4 g^2 {\nu}^2
(ys)
v^m     s_k   r^i{}_m
+ 4 g^2 {\nu}^2
(ys)
v^m     s_m   r^i{}_k
$$

\ses

$$
-   g^2   {\nu}
(ys)
v^m     (ys)(\nu_m  r^i{}_k+\nu_k  r^i{}_m)
$$

\ses

            \ses

    \ses

$$
+
\nu_j
 g^2
(ys)
\Bigl[
-
\nu_k
(ys)
v^j
+        2   {\nu}   s_k
v^j
+
\nu
(ys)
r^j{}_k
\Bigr]
v^i
$$

\ses

    \ses

$$
+        2  g^2  {\nu}   s_j
\Bigl[
\nu_k
(ys)
v^j
-        2  {\nu}   s_k
v^j
-
{\nu}
(ys)
r^j{}_k
\Bigr]
v^i
$$

\ses

    \ses

$$
+
  g^2  {\nu}
(ys)
r^i{}_j
\Bigl[
\nu_k
(ys)
v^j
-        2   {\nu}   s_k
v^j
-
{\nu}
(ys)
r^j{}_k
\Bigr]
+\De,
$$

{

\ses
\nin
or, on reducing similar terms,
$$
2\nu^4G^jG^i{}_{kj}
-
\nu^4
G^i{}_jG^j{}_k
=
$$

\ses

\ses

$$
2  g^2
(ys)
v^m\Biggl(
2\nu_k\nu_m-{\nu}\fr1q\eta_{km}
\Biggr)
(ys)
v^i
+  4  g^2
(ys)
v^m   {\nu}^2   (\nabla_m b_k)   v^i
$$

\ses

$$
- 2  g^2   \nu
(ys)
v^m
   (\nu_m  s_k +\nu_ks_m)  v^i
+ 2 g^2 {\nu}^2
(ys)
v^m     s_k   r^i{}_m
+ 4 g^2 {\nu}^2
(ys)
v^m     s_m   r^i{}_k
$$

\ses

$$
-   g^2   {\nu}
(ys)
v^m     (ys)\nu_m  r^i{}_k
$$

\ses

            \ses

    \ses

$$
+
\nu_j
 g^2
(ys)
\Bigl[
-
\nu_k
(ys)
v^j
+
\nu
(ys)
r^j{}_k
\Bigr]
v^i
$$

\ses

    \ses

$$
+        2  g^2  {\nu}   s_j
\Bigl[
-        2  {\nu}   s_k
v^j
-
{\nu}
(ys)
r^j{}_k
\Bigr]
v^i
$$

\ses

    \ses

\be
-
  g^2  {\nu}
(ys)
r^i{}_j
{\nu}
(ys)
r^j{}_k
+\De.
\ee

\ses

\ses

On applying the contractions
\be
v^ms_m=(ys)-b \si,
\qquad
v^jr^i{}_j= v^i  -  (1-c^2)  b  b^i,
\ee
\ses
\be
r^j{}_mr^i{}_j= r^i{}_m -          (1-c^2)b^ib_m,
\qquad
v_jv^j=   q^2-(1-c^2)       b^2,
\ee
\ses
and
\be
\nu_jv^j= \nu -(1-c^2)\fr1q(b^2+gc^2   bq)
\ee
(see [1]),
{
the above expression  reads
$$
2\nu^4G^jG^i{}_{kj}
-
\nu^4
G^i{}_jG^j{}_k
=
4  g^2
(ys)
\nu_k
\Bigl[\nu -(1-c^2)\fr1q(b^2+gc^2   bq)\Bigr]
(ys)
v^i
$$

\ses

\ses

$$
-2  g^2
(ys){\nu}\fr1q\Bigl[
v_k-(1-c^2)bb_k-\fr1{q^2}v_k
(q^2-(1-c^2)b^2 )
\Bigr]
(ys)
v^i
$$

\ses

$$
+  4  g^2
(ys)
 {\nu}^2 (  s_k -b\si_k)  v^i
$$

\ses

$$
- 2  g^2   \nu
(ys)
\Biggl(\Bigl[\nu -(1-c^2)\fr1q(b^2+gc^2   bq)\Bigr]  s_k
 +\nu_k\Bigl((ys)-b\si\Bigr)
 \Biggr)  v^i
$$

\ses

$$
+ 2 g^2 {\nu}^2
(ys)
 s_k
\bigl[v^i  -  (1-c^2)  b  b^i\bigr]
$$

\ses

\ses

$$
+ 4 g^2 {\nu}^2
(ys)
   \bigl((ys)-b\si\bigr)
   r^i{}_k
$$

\ses

$$
-   g^2   {\nu}
(ys)
     (ys)
\Bigl[\nu -(1-c^2)\fr1q(b^2+gc^2   bq)\Bigr]
      r^i{}_k
$$

\ses

            \ses

    \ses

$$
+
\nu_j
 g^2
(ys)
(ys)
({\nu}r^j{}_k-\nu_kv^j)
v^i
-        2  g^2  {\nu}   s_j
\Bigl[(ys)  r^j{}_k+2 s_kv^j\Bigr]
v^i
$$

\ses

    \ses

\be
-
  g^2  {\nu}^2
(ys)
(ys)
\Bigl[ r^i{}_k -          (1-c^2)b^ib_k  \Bigr]
+\De.
\ee

{

\ses

For   the $hh$-curvature tensor $R^i{}_k$ we may use the formula
$$
K^2R^i{}_k~:=
2\D{\bar G^i}{x^k}-\D{\bar G^i}{y^j}\D{\bar G^j}{y^k}
-y^j\Dd{\bar G^i}{x^j}{y^k}
+2\bar G^j\Dd{\bar G^i}{y^k}{y^j}
$$
\ses
\be
=
2\D{\bar G^i}{x^k}-\bar G^i{}_j\bar G^j{}_k
-y^j\D{\bar G^i{}_k}{x^j}
+2\bar G^j\bar G^i{}_{kj}
\ee
(which is tantamount to the definition (3.8.7) on p. 66 of the book [7]);
we apply the notation
$$
 \bar G^i=\fr12 G^i, \qquad \bar  G^i{}_k=\fr12  G^i{}_k, \qquad  \bar G^i{}_{km}=\fr12 G^i{}_{km},
\qquad  \bar G^i{}_{kmn}=\fr12 G^i{}_{kmn}.
$$

\ses

\vskip 1cm

\def\bibit[#1]#2\par{\rm\noindent\parskip1pt
                     \parbox[t]{.05\textwidth}{\mbox{}\hfill[#1]}\hfill
                     \parbox[t]{.925\textwidth}{\baselineskip11pt#2}\par}

\nin
{  REFERENCES}

\ses

\ses

\bibit[1] G. S. Asanov:   Finsleroid-regular    space  developed.    Berwald case, {\it  arXiv:math.DG}/0711.4180v1 (2007).

\ses

\bibit[2] G. S. Asanov:  Finsleroid--Finsler  space with Berwald and  Landsberg conditions,
 {\it  arXiv:math.DG}/0603472 (2006).

\ses

\bibit[3] G. S. Asanov:  Finsleroid--Finsler  space and spray   coefficients, {\it  arXiv:math.DG}/0604526 (2006).

\ses

 \bibit[4] G. S. Asanov:  Finsleroid--Finsler  spaces of positive--definite and  relativistic types.
 \it Rep. Math. Phys. \bf 58 \rm(2006), 275--300.

\ses

\bibit[5] G. S. Asanov:  Finsleroid--Finsler space and geodesic spray    coefficients,
{\it Publ.  Math. Debrecen } {\bf 71/3-4} (2007), 397-412.

\ses

\bibit[6]   G.S. Asanov:
Finsleroid  corrects     pressure  and energy   of
 universe. Respective  cosmological equations, {\it  arXiv}:0707.3305 [math-ph] (2007).

\ses

 \bibit[7] D. Bao, S. S. Chern, and Z. Shen: {\it  An
Introduction to Riemann-Finsler Geometry,}  Springer, N.Y., Berlin (2000).





\ses

\bibit[8] J.L. Synge: \it Relativity: the General Theory, \rm North--Holland, Amsterdam  (1960).

\ses

\bibit[9] S. Weinberg: \it Gravitation and Cosmology, \rm John Wiley, New York (1972).

\ses

\end{document}